\documentclass[sigconf]{acmart}
\usepackage{booktabs} 
\usepackage{fancyhdr}
\usepackage{enumitem}
\usepackage{subcaption}
\usepackage{multirow}
\usepackage{listings}
\usepackage{xcolor}
\usepackage{amsmath}

\usepackage{mathrsfs,amssymb}
\usepackage[boxed,ruled]{algorithm2e}
\SetKwRepeat{Do}{do}{while}
\usepackage{makecell}
\usepackage[para,online,flushleft]{threeparttable}
\usepackage{mathtools}
\usepackage{bm} 
\DeclarePairedDelimiter{\ceil}{\lceil}{\rceil}
\DeclareMathAlphabet{\mathcal}{OMS}{cmsy}{m}{n}

\newcommand{\project}{{\sc Ftrans}\xspace}

\copyrightyear{2020}
\acmYear{2020}
\setcopyright{acmcopyright}\acmConference[ISLPED '20]{ACM/IEEE International Symposium on Low Power Electronics and Design}{August 10--12, 2020}{Boston, MA, USA}
\acmBooktitle{ACM/IEEE International Symposium on Low Power Electronics and Design (ISLPED '20), August 10--12, 2020, Boston, MA, USA}
\acmPrice{15.00}
\acmDOI{10.1145/3370748.3406567}
\acmISBN{978-1-4503-7053-0/20/08}

\begin{document}

\title{FTRANS: Energy-Efficient Acceleration of Transformers using FPGA}

\author{Bingbing Li$^1$, Santosh Pandey$^2$, Haowen Fang$^3$, Yanjun Lyv$^1$, Ji Li$^4$, Jieyang Chen$^5$, Mimi Xie$^6$, Lipeng Wan$^5$, Hang Liu$^2$ and Caiwen Ding$^1$}

\affiliation{\small $^1$University of Connecticut 
\quad $^{2}$Stevens Institute of Technology
\quad $^{3}$Syracuse University \\
$^{4}$Microsoft Corporation 
\quad $^{5}$Oak Ridge National Laboratory
\quad $^{6}$University of Texas at San Antonio\\
$^1$\{bingbing.li, lyu.yanjun, caiwen.ding\}@uconn.edu  \quad 
$^2$\{spande1, Hang.liu\}@stevens.edu \quad 
$^3$hfang02@syr.edu \\
$^4$changzhouliji@gmail.com \quad 
$^5$\{chenj3, wanl\}@ornl.gov \quad 
$^6$mimi.xie@utsa.edu
}

\renewcommand{\shortauthors}{Bingbing Li et al.}

\begin{abstract}
In natural language processing (NLP), the ``Transformer" architecture was proposed as the first transduction model replying entirely on self-attention mechanisms without using sequence-aligned recurrent neural networks (RNNs) or convolution, and it achieved significant improvements for sequence to sequence tasks. The introduced intensive computation and storage of these pre-trained language representations has impeded their  popularity  into  computation and  memory constrained devices. The field-programmable gate array (FPGA) is widely used to accelerate deep learning algorithms for its high parallelism and low latency. However, the trained models are still too large to accommodate to an FPGA fabric.
In this paper, we propose an efficient acceleration framework, \project, for transformer-based large scale language representations. Our framework includes enhanced block-circulant  matrix (BCM)-based weight representation to enable model compression on large-scale language representations at the algorithm level with few accuracy degradation, and an acceleration design at the architecture level. Experimental results show that our proposed framework significantly reduce the model size of NLP models by up to 16 times. Our FPGA design achieves 27.07$\times$ and 81 $\times$ improvement in performance and  energy efficiency compared to CPU, and up to 8.80$\times$ improvement in energy efficiency compared to GPU.

\end{abstract}

\maketitle

\section{Introduction}

RNN and its variant \emph{Long Short-Term Memory} (LSTM) unit \cite{hochreiter1997long} and \emph{Gated Recurrent unit} (GRU) \cite{cho2014properties} used to dominate in sequence modeling, language modeling and machine translation, etc.
However, they in general lack efficiency in transmitting global information, due to the bottleneck in the memory (hidden state) and complicated bypassing logic (additive and derivative branches) where long range information is passed. 
In addition, the inherently sequential nature precludes parallelization within training examples through backpropagation, which is critical at longer sequence lengths~\cite{kim2017structured}. 

To overcome the shortcomings in RNNs, the ``Transformer" architecture was proposed as the first transduction model replying entirely on self-attention mechanisms without using sequence-aligned RNNs or convolution. It achieved notable improvements for sequence to sequence tasks \cite{vaswani2017attention}. 
The breakthroughs and developments of new models have accelerated at an unprecedented pace since the attention mechanisms have become the mainstream in NLP domain with the invention of Transformer. 
Many transformer-based NLP language models like BERT~\cite{devlin2018bert} and RoBERTa~\cite{liu2019roberta} introduced pretraining procedures to the transformer architecture and achieved record-breaking results on major NLP tasks, including question answering, sentiment analysis, and language inference.

Nevertheless, the introduced intensive computation and power footprint of these pre-trained language representations has impeded their popularity into computation and energy constrained as edge devices. 
Moreover, despite of the rapid advancement achieved by the recent transformer-based NLP models, there is a serious lack of studies on compressing these models for embedded and internet-of-things (IoT) devices. 

In this paper, we propose an energy-efficient acceleration framework, \project, for transformer-based large scale language representations using FPGA. \project is comprised of an enhanced BCM-based method enabling model compression on 
language representations at the algorithm level, and an acceleration design at the architecture level. 
Our contributions are summarized as follows:
\leftmargini=4mm
\begin{itemize}

\item \textbf{Enhanced BCM-based model compression for Transformer.}
We address the accuracy degradation caused by traditional BCM compression, and propose an enhanced BCM-based compression to reduce the footprint of weights in Transformer. With small accuracy loss, \project achieves up to 16 times compression ratio.

\item \textbf{Holistic optimization for Transformers on FPGA.}
Given the large size and complex data flow of transformer-based models, even with model compression, we still need to schedule the computation resources carefully to optimize latency and throughput.
We propose a two stage optimization approach to mitigate the resource constraints and achieve high throughput.

\item \textbf{Low hardware footprint and low power (energy) consumption. }
We propose an FPGA architecture design to support the model compression technique and we develop a design automation and optimization technique. Overall, the proposed \project achieves the lowest hardware cost and energy consumption in implementing Transformer and RoBERTa compared to CPU and GPU references.

\end{itemize}
Experimental results show that our proposed framework significantly reduce the size of NLP models by up to 16 times. Our FPGA design achieves 27.07$\times$ and 81 $\times$ improvement in performance and  energy efficiency compared to CPU. The power consumption of GPU is up to 5.01$\times$ compared to that of FPGA, and we achieve up to 8.80$\times$ improvement in energy efficiency compared to GPU.

\section{Related Work}

Attention mechanisms have become an integral part of compelling sequence modeling and transduction models in various tasks
\cite{kim2017structured}.
Evidence of NLP community moving towards attention-based models can be found by more attention-based neural networks developed by companies like Amazon \cite{kim2018supervised}, Facebook \cite{shi2018simple}, and Salesforce \cite{bradbury2017towards}. 
The novel approach of Transformer is the first model to eliminate recurrence completely with self-attention to handle the dependencies between input and output. 
BERT~\cite{devlin2018bert} and RoBERTa~\cite{liu2019roberta} extend Transformer's capacity from a sequence to sequence model to a general language model by introducing the pretraining procedure, and achieved state-of-the-art results on major NLP benchmarks.
Although RNNs and Convolutional Neural Networks (CNNs) are being replaced by Transformer-based models in NLP community, there are only a few works that accelerate Transformers and focus on reducing the energy and power footprint, e.g., a case study of Transformer is presented in \cite{xilinx} using one of the cutting-edge FPGA boards. However, it is noteworthy that \cite{xilinx} targets at a specialized FPGA architecture, which employs High Bandwidth Memory (HBM) technology. Unlike the conventional FPGA, HBM is packaged directly within the FPGA fabric to alleviate the on chip memory constraint. 
However, work \cite{xilinx} did not adopt model compression technique, and used the sequence length of 8 and 16, which are too short and not favorable in practise. The model details such as number of encoder/decoders, hidden size are also not listed.




\section{Transformer Workload Analysis}
\begin{figure} [b]
     \centering
     \includegraphics[width=0.9\columnwidth]{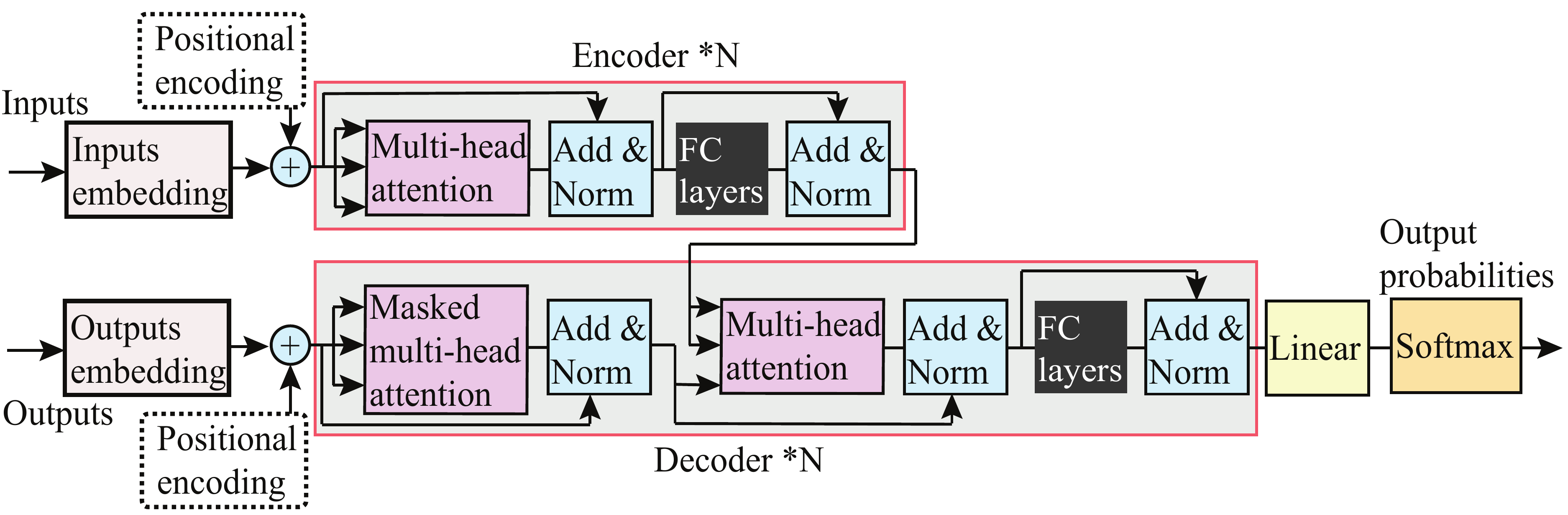}     \caption{Model structure of Transformer.}
     \label{fig:transformer}
\end{figure}

The ``Transformer" architecture is the heart for all state-of-the-art large scale language models. It has 
an encoder-decoder structure~\cite{vaswani2017attention} as shown in Figure~\ref{fig:transformer}. The encoder maps a sequence of the input symbols $\mathbf{x}= ({x}_1; {x}_2; {x}_3; ...; {x}_n)$ to a sequence of continuous representations $\mathbf{z}= ({z}_1; {z}_2; {z}_3; ...; {z}_n)$. Given $\mathbf{x}$, the decoder then produces an output
sequence $\mathbf{y}= ({y}_1; {y}_2; {y}_3; ...; {y}_m)$ of symbols one element per time step. For the next time step, the model takes the previously generated symbols as additional input when generating the next.
The Transformer follows this overall architecture using stacked self-attention and fully-connected (FC) layers for both the encoder and decoder, shown in Figure~\ref{fig:transformer}.

\textbf{Encoder:} The encoder consists of a stack of $N$
identical layers. Each layer has two
sub-layers. The first is a multi-head self-attention mechanism, and the second is a FC feed-forward network. 
There is a residual connection  around each of the two sub-layers, followed by layer normalization.

\textbf{Decoder:} The decoder contains of a stack of $N$ identical layers. Within each layer, there are three sub-layers, where the third sub-layer is the same as the encoder. The inserted second sub-layer
performs multi-head attention over the output of encoder stack. The first-sublayer utilizes masked multi-head attention, 
to ensure that predictions for position $i$ only depends on its previous positions.


\subsection{Attention}
\begin{figure} [b]
     \centering
     
     \includegraphics[width=0.5\columnwidth]{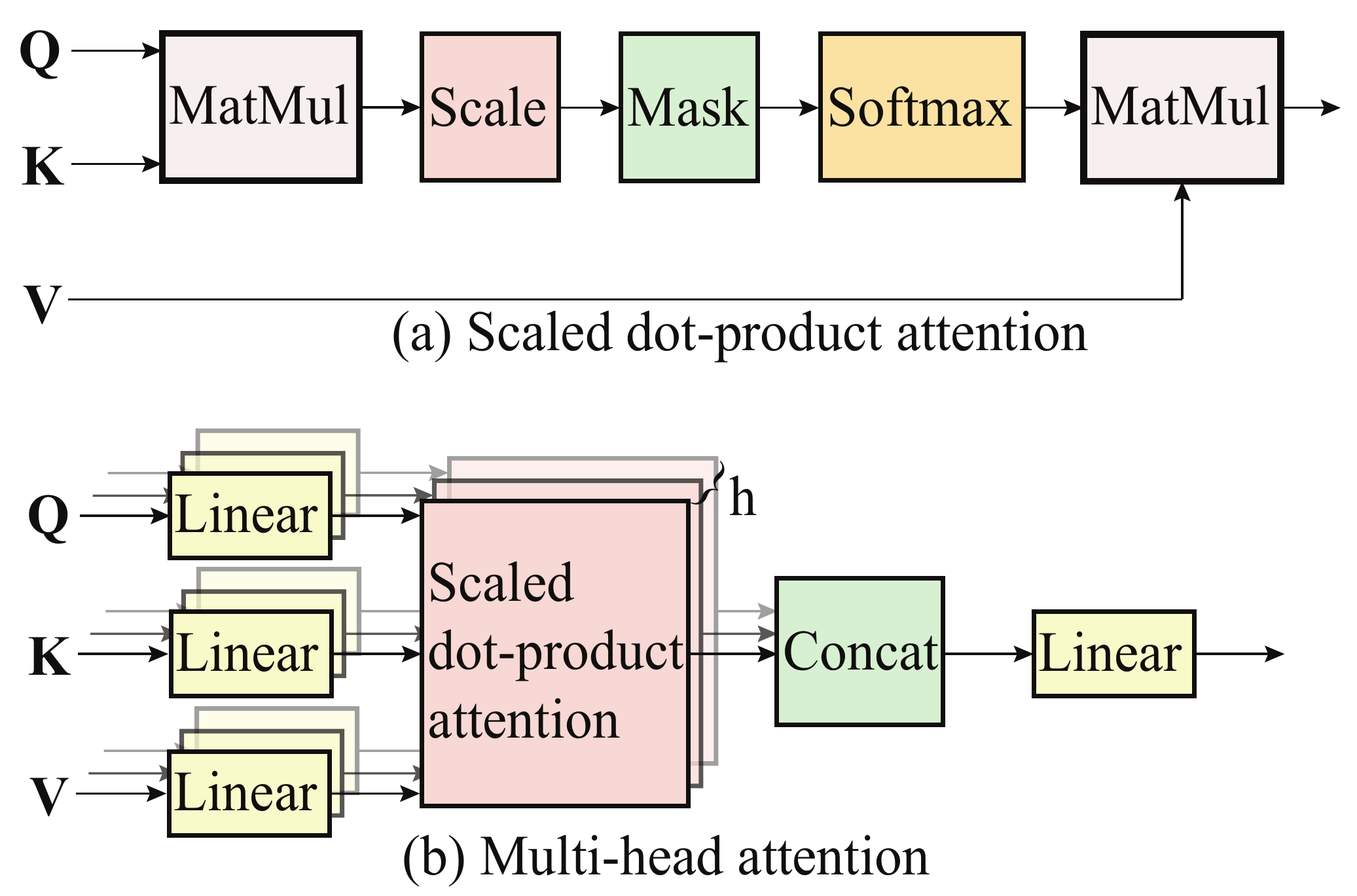}     \caption{: (a) Scaled Dot-Product Attention. (b) Multi-Head Attention.}
     \label{fig:attention}
\end{figure}

The attention function can be described as mapping a query $\mathbf{q}$ and a set of keys $\mathbf{k}$ and values $\mathbf{v}$ pairs to an output $\mathbf{o}$ as shown in Figure~\ref{fig:attention} (a), named scaled dot-product attention, or single head attention. 

\subsubsection{Single Head Attention} In this paper, we select dot-product attention as the attention function since it is much faster and more space-efficient ~\cite{vaswani2017attention}. 
The input consists of
queries and keys of dimension $d_k$, and values of dimension $d_v$. We denote $\sqrt{d_k}$ is the scaling factor for dot-product attention. We compute the dot products of the
query with all keys, divide each by $\sqrt{d_k}$, and apply a softmax function to obtain the weights on the
values. The attention function on $\mathbf{q}$, $\mathbf{k}$, and $\mathbf{v}$ can be computed simultaneously by concatenated 
into matrix $\mathbf{Q}$, $\mathbf{K}$, and $\mathbf{V}$, respectively. Accordingly, the output matrix $\mathbf{O}_{att}$ is:

\vspace{-0.2cm}
\begin{equation}
\small
\label{eqn:singlehead}
\mathbf{O}_{att} = Attention(\mathbf{Q}, \mathbf{K}, \mathbf{V} ) = softmax(\frac{\mathbf{QK}^T}{\sqrt{d_k}})
\vspace{-0.2cm}
\end{equation}

\subsection{Multi-head Attention}

Multi single-head attention are then concatenated as multi-head attention, as shown in Figure~\ref{fig:attention} (b). 
MultiHead ($\mathbf{Q}$, $\mathbf{K}$, $\mathbf{V}$) = Concat ($\text{Head}_1, \cdots, \text{Head}_h$)$\times\mathbf{W}^O$, where the Head is defined as:
\vspace{-0.2cm}
\begin{equation}
\small
\label{eqn:attention}
\text{Head}_i = Attention(\mathbf{Q}\mathbf{W}_i^Q, \mathbf{K}\mathbf{W}_i^K, \mathbf{V}\mathbf{W}_i^V)
\vspace{-0.2cm}
\end{equation}
where the projections are parameter matrices ${{\bf{W}}^Q_i}\in\mathbb{R}^{d_{model} \times d_k}$, ${{\bf{W}}^K_i}\in\mathbb{R}^{d_{model} \times d_k}$, and ${{\bf{W}}^V_i}\in\mathbb{R}^{d_{model} \times d_v}$. Multi-head attention enables the model to jointly attend to information from different representation subspaces at different positions~\cite{vaswani2017attention}. 

{In this work, we implement a shallow Transformer and a large scale Transformer, i.e., RoBERTa. The shallow Transformer has $h$ = 2 parallel attention layers with 4 attention heads and RoBERTa (base configuration) has 12 layers with 12 heads. For each head we use
$d_k = d_v = d_{model}/h = 200$ and $768$ for Transformer and RoBERTa, respectively. 
}


\section{Transformer Compression using Enhanced Block-Circulant Matrix}

The introduced intensive computation and weight storage of large pre-trained language representations have brought challenges in hardware implementation. Therefore, model compression is a natural method to mitigate the these challenges. 


\subsection{Enhanced BCM-based Transformer}
\label{sec:structured}

CirCNN~\cite{ding2017circnn} and C-LSTM~\cite{Wang2018clstm} have adopted BCM for model compression on small to medium scale datasets in image classification and speech recognition, respectively, and achieved significant improvement in terms of performance and energy efficiency compared to the prior arts. 
Using this method, we can reduce weight storage by replacing the original weight matrix with one or multiple blocks of circulant matrices, where each row/column is the cyclic reformulation of the others.
We use $b$ to represent the row/column size of each circulant matrix (or block size, FFT size). Suppose the shape of a weight matrix in Transformer (e.g., ${{\bf{W}}^Q_i},{{\bf{W}}^K_i},{{\bf{W}}^V_i})$ is $\textbf{W}\in \mathbb{R}^{m\times n}$, there will be $f \times g$ blocks after partitioning $\mathbf{W}$, where $f = m\div b$ and $g=n \div b$. Then $\mathbf{W} = [\mathbf{W}_{ij}]$, $i \in \{1 \dots f\}$, $j \in \{1 \dots g\}$.

The input $\mathbf{x}$ is also partitioned as $\mathbf{x} = [\mathbf{x}^T_1, \mathbf{x}^T_2, \dots, \mathbf{x}^T_g]^T$. In each BCM, only the first column/row is needed for storage and computation, and is termed the \emph{index vector}, $\mathbf{p}_{ij}$. 
The theoretical foundation is derived in \cite{zhao2017theoretical}, which demonstrates the universal approximation property and the error bounds of BCM-based neural networks are as efficient as general neural networks.


Prior works~\cite{ding2017circnn, Wang2018clstm} have not investigated large-scale language representations. To further maintain the prediction accuracy, we use an enhanced BCM-based model compression. We modify the formulation of the index vector as follows:
\begin{equation}
\footnotesize
\mathbf{p}_{ij} 
=
\begin{bmatrix}
         \frac{1}{b}\sum_{j=1}^b \mathbf{W}_{1j}    \\
         \frac{1}{b}\sum_{j=1}^b \mathbf{W}_{2j}    \\
         \dots \\
         \frac{1}{b}\sum_{j=1}^b \mathbf{W}_{bj}  
\end{bmatrix}
\end{equation}
where $\mathbf{W}_{ij}$ is a circulant matrix. 
We observe that in this way, we can better preserve the parameter information and maintain the overall prediction accuracy. The main reason is that prior works take the first column/row as the \emph{index vector}, missing the effective representations for other rows/columns.

Based on the \emph{circulant convolution theorem} \cite{pan2012structured,smith2007mathematics}, instead of directly performing the matrix-vector multiplication, we could use the fast Fourier transform (FFT)-based multiplication method, and it is equivalent to matrix-vector multiplication. The calculation of a BCM-based matrix-vector multiplication $\mathbf{W}_{ij} \mathbf{x}_j$ is:
$\mathbf{W}_{ij}\mathbf{x}_j=\mathbf{p}_{ij}\circledast \mathbf{x}_j=\text{IFFT}\big(\text{FFT}(\mathbf{p}_{ij})\circ\text{FFT}(\mathbf{x}_j)\big)$,
where `$ \circledast$ ' represents circular convolution, and $\circ$ is element-wise multiplication. Therefore, the {computational complexity is reduced from O($b^2$) to O($b\log b$)}.


\section{Architecture}

FPGA is widely used to accelerate deep learning models for its high parallelism and low latency. 
As large amount of transformer parameters exceed the on-chip memory or block RAM (BRAM) capacity on FPGA fabric, even with model compression technique, the full model cannot be stored on chip.
To address the challenge, we partition a model into embedding layer and encoder/decoder stacks. The embedding layer contributes 30.89\% of parameters.
Essentially it is a look-up table which transforms discrete tokens into continuous space, 
the computation is less than that of encoder and decoder. Therefore, our basic idea is to off-load embedding layer to off-chip memory, thus it is possible to deploy the most computational intensive part, i.e. the encoder and decoder stack on chip, avoiding frequently access off-chip weights, hence to accelerate computation. Second, to mitigate the I/O constrain, we developed the inter-layer coarse grained pipelining, intra-layer fine grained pipeling, and computation scheduling.
\subsection{Overall Hardware Architecture}

 
  \begin{figure}[b]
    \centering
    \includegraphics[width=0.8\columnwidth]{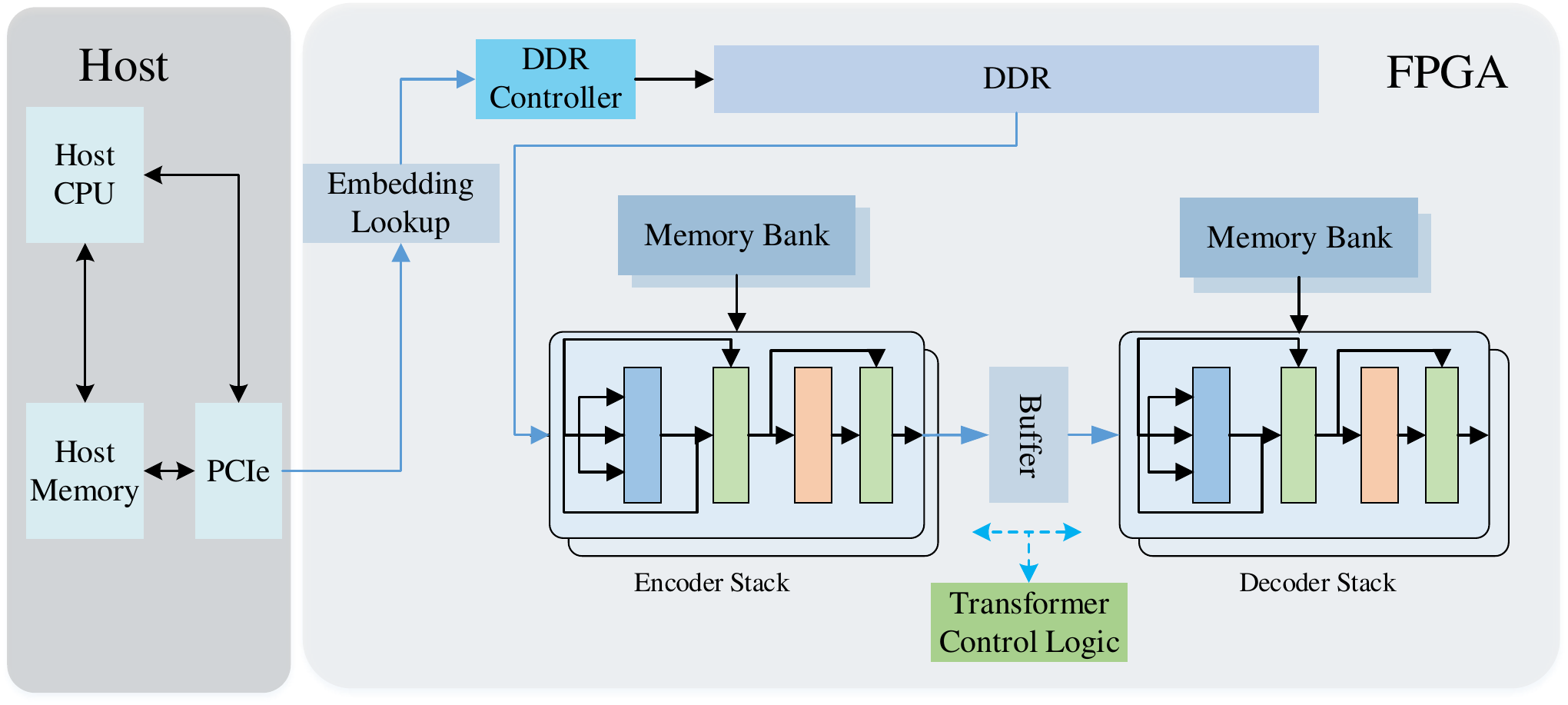} 
    \vskip -0.5em
    \caption{The overall hardware architecture on FPGA.} 
    \label{fig:sys_arch}
 \end{figure}

As shown in Figure~\ref{fig:sys_arch}, the proposed hardware architecture consists of computation units for encode/decoder computation, on-chip memory banks, a transformer controller, and an off-chip memory (DDR) and DDR controller. The transformer controller communicates with the host and controls all the modules in FPGA. The host PC loads the inputs (i.e., sentence pairs ) to the FPGA  for inference through PCIE. On the FPGA part, given the tokenized sentences, the embedding look up module accesses DDR to fetch embeddings. Next, the embeddings will be fed into the pipelined encoder/decoder stacks to perform inference.

The computing units consist of multi-head attention, scaled dot product attention, point wise feed forward layer, linear, and add/norm. The transformer controller orchestrates the computing flow and data flow of inputs from PCIEs, BRAMs and computing units on the FPGA fabric. Since the encoder and decoder share same type of operations, so we first decompose them into different computing primitives, including matrix multiplication of different sizes, vectorized exponentials etc. The multi-head attention, linear, and add/norm modules are reconfigured to form as encoder or decoder under the transformer control logic. We have two pipeline strategies. For shallow networks, the entire network can be straightforwardly implemented, i.e. all layers can be implemented by dedicated FPGA resources. For the state-of-the-art designs such as BERT and RoBERTa, there are multiple encoders/decoders, hence the resource such as DSPs may not enough. In such cases, reuse of certain PE or entire encoder/decoder module are necessary.

\begin{figure} [t]
     \centering
     \includegraphics[width=0.65\columnwidth]{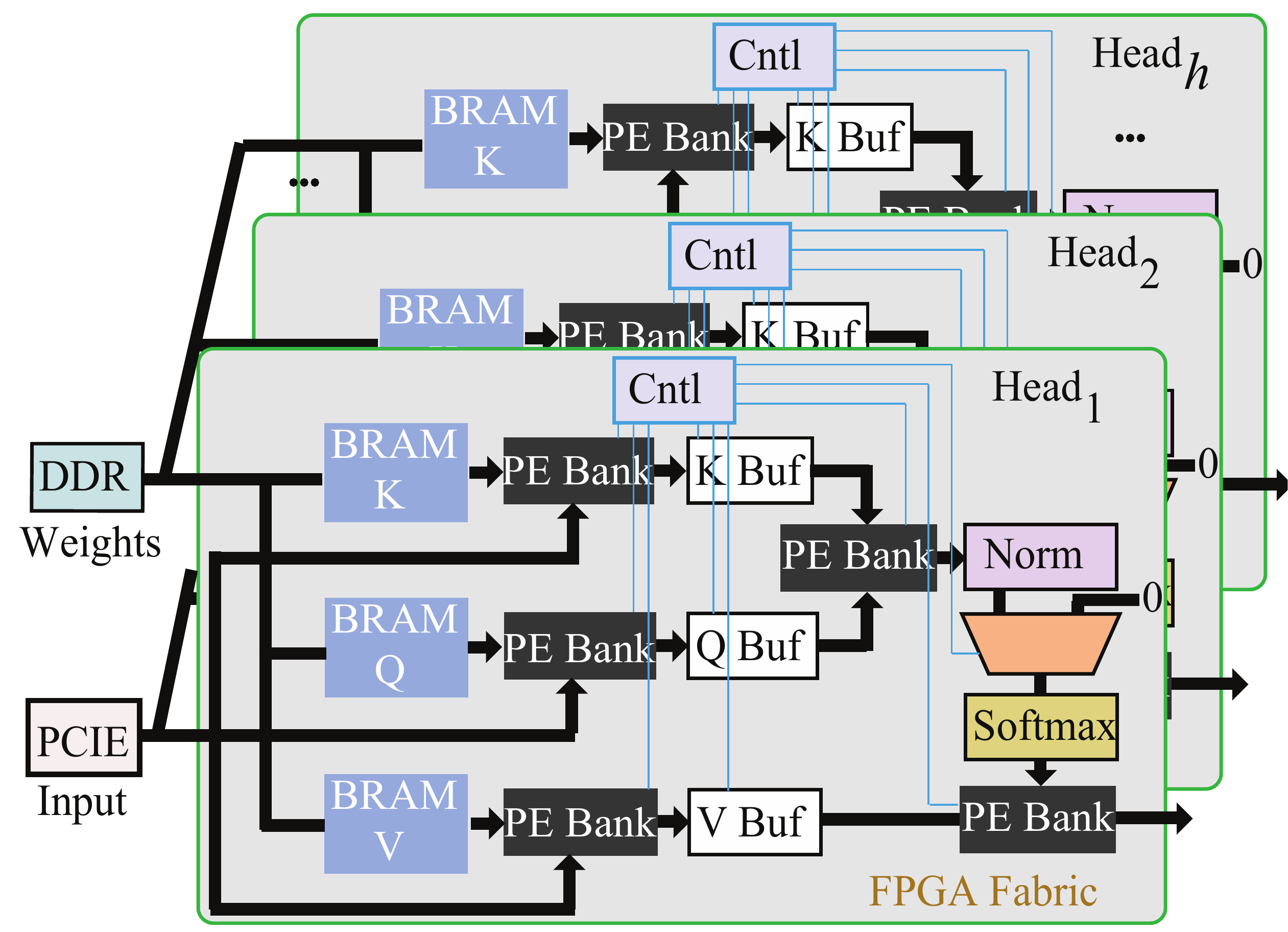}     \caption{Multi-head attention ($\text{Head}_1, \cdots, \text{Head}_h$) design.
     }
     \label{fig:multihead}
\end{figure}

\subsection{Multi-Head Attention Design}

  \begin{figure}[b]
    \centering
    \includegraphics[width=0.7\columnwidth]{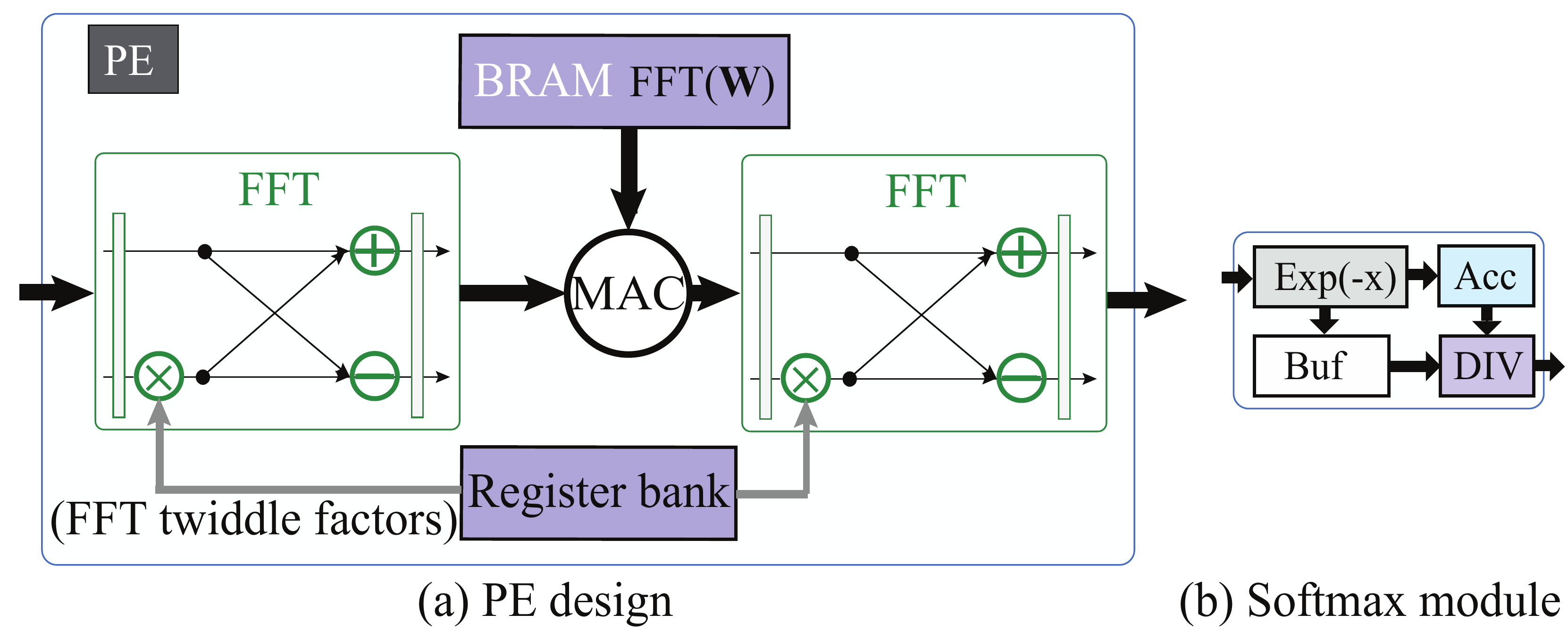} 
    \vskip -0.5em
    \caption{FFT/IFFT-based PE and softmax module design.}
    \label{fig:PE} 
 \end{figure}
 Multi-head attention includes multi-\emph{processing elements} (named PE) banks, for matrix multiplication), buffers (K buf, Q buf, and V buf), a normalization module (Norm), a masking function for masked  multi-head attention, and a softmax module as described in Equation (\ref{eqn:attention}) and shown in Fig.~\ref{fig:multihead}.

The input are fetched from DDR and fed into encoder pipeline, then multiplied with a set of query matrix $\mathbf{Q}$ and key matrix $\mathbf{K}$ stored on BRAMs. The intermediate results $\mathbf{Q}\mathbf{W}^Q$ and $\mathbf{K}\mathbf{W}^K$ are then propagated to the buffers (i.e., K buffer and Q buffer to store $\mathbf{K}\mathbf{W}^K$, and $\mathbf{Q}\mathbf{W}^Q$, respectively). Next, we compute the matrix multiplication of the values stored in the K buffer and W buffer. The product will be loaded to the normalization (Norm) module, i.e.,$\frac{product}{\sqrt{d_k}}$.
After the softmax module, the results will be propagated to a PE bank to perform matrix multiplication with the matrix stored in $V$ Buf, i.e., $\mathbf{V}\mathbf{W}^V$. Each head will have a local controller to orchestrate the computing flow and data flow of PE banks and buffers. The local controller will also enable the multiplexer for masked multi-head attention with a goal of masking the future tokens in a sequence (setting to 0), to prevent the current output predictions from being able to see later into the sentence. To support masked multi-head attention, the local controller controls multiplexer to set future tokens to 0, such that current output predictions are not able to see later sequences. Decoder has one more multi-head attention, thus takes longer time to compute than encoder. In the case of decoder module has to be reused, to prevent encoder pipeline stall, a buffer is placed between the encoder and decoder stacks. The buffer also stores the output from the last encoder to support the residue connection.



\subsection{PE Design and Softmax Module}
We develop three different configurable PEs, which as PE-A, PE-B, and PE-FFT/IFFT. 
For the BCM-based matrix-vector multiplication in FC layers, we use FFT/IFFT-based processing elements (PE); for other layers, we use matrix-vector multiplication, i.e., PE-A and PE-B for different matrix sizes.

\subsubsection{Matrix-vector multiplication-based PE} 
The major part of the PE-A or PE-B is a multiplier for matrix multiplication of different sizes. It also consists of two accumulators, dividers and exponential units to support scaling and softmax required by multi-head attention. The output of multipliers are fed into divider or accumulator as stream, hence scaling and softmax layer can be overlapped with matrix multiplication.

\subsubsection{FFT/IFFT-based PE }
Figure~\ref{fig:PE} shows the design of FFT/IFFT-based PE and softmax, including a FFT/IFFT kernel, an accumulator, and an adder.
The accumulator is an adder tree with $N$ inputs (the size is chosen the same as the FFT/IFFT kernel size). 
We select Radix-2 Cooley Tukey algorithm~\cite{johnsson1992cooley} for FFT implementation.

\subsubsection{Softmax Module}
Figure~\ref{fig:PE} (b) shows the implementation of the softmax function $\text{softmax}(x)_i = \frac{exp(x_i)}{\sum_{j}^{ }exp(x_j))}$. The exponential function ${exp(x_i)}$ or ${exp(x_j)}$ is expensive in resource consumption for FPGAs. We adopt piece-wise linear functions to estimate their outputs, in order to simultaneously reduce the resource consumption and maintain the accuracy. A buffer is used to store ${exp(x_i)}$ and an accumulator is used to compute the summation of ${exp(x_j)}$. Next, we perform the division and generate the softmax results.

\section{Design Automation \& Optimization}

We developed a workflow to prototype and explore the hardware architecture. First, we generate a data dependency graph based on trained models to illustrate the computation flow. The operators in graph are scheduled to compose the pipeline under the design constraints, to achieve maximum throughput. At last, a code generator receives the scheduling results and generates the final C/C++ implementation, which can be fed into the commercial HLS tool for synthesis. Our target synthesis backend is Xilinx SDx.

\begin{figure} [b]
     \centering
     \includegraphics[width=0.8\columnwidth]{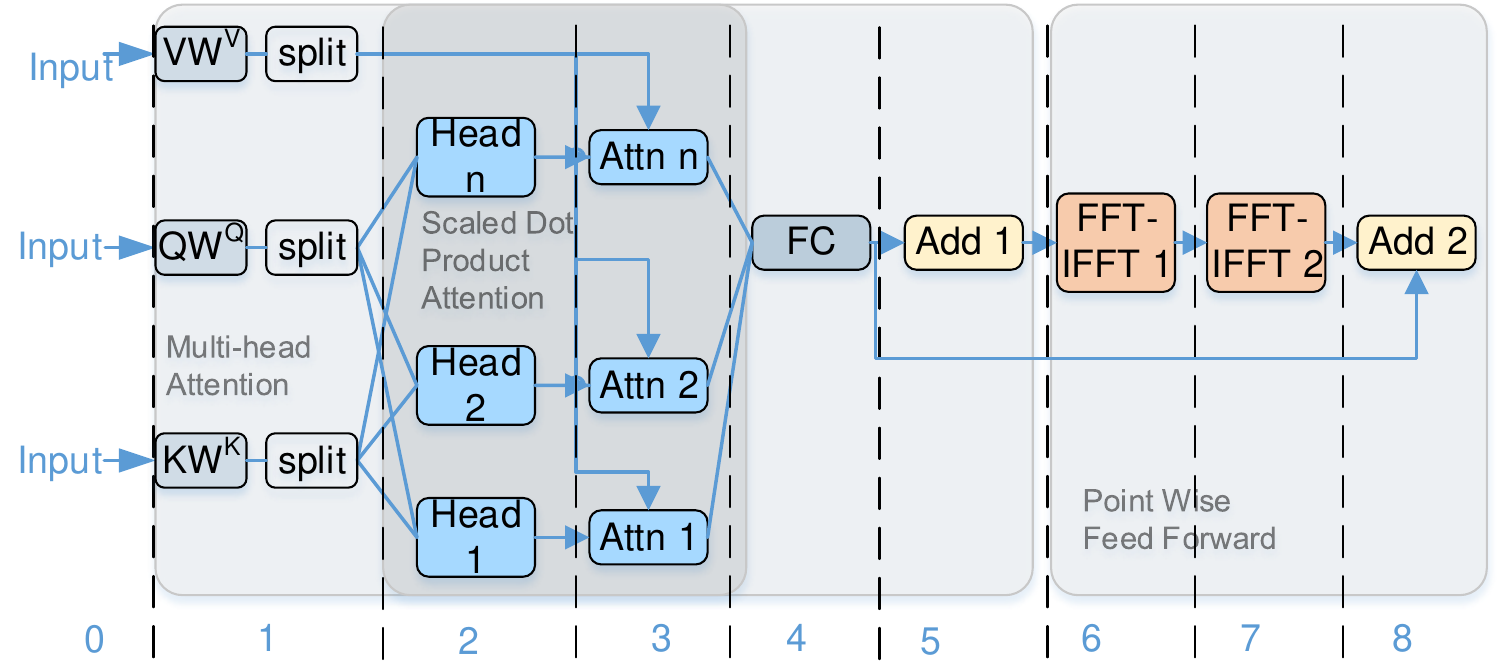}\caption{Data flow of major operations.}
     \label{fig:flow}
\end{figure}

\begin{figure} [b]
     \centering
     \includegraphics[width=0.8\columnwidth]{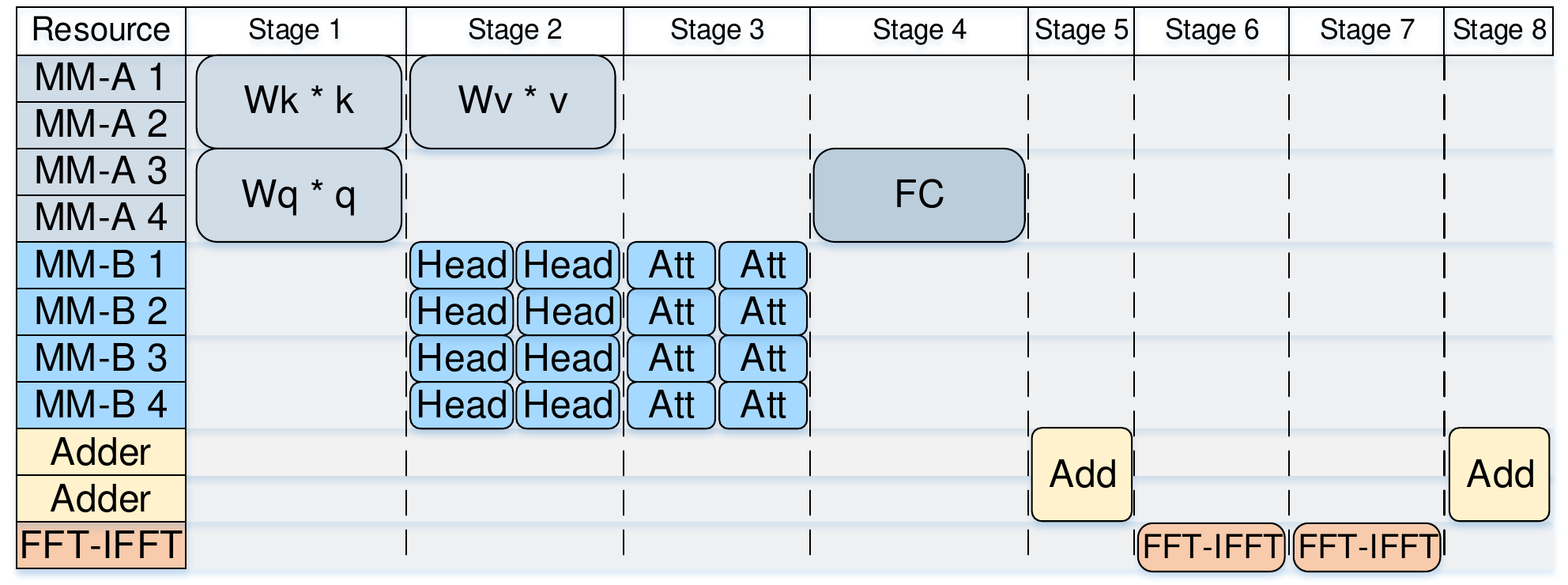}\caption{Fine grained  operation scheduling}
     \label{fig:pipeline}
\end{figure}

The major computationally intensive operations are shown in Figure \ref{fig:flow}. Other operations such as division and softmax consume much less time, and can be merged/overlapped with these major operations. The computation in different layers can be decomposed into common computing elements, i.e., PEs. The layers in same color can be performed by same PE, however, with unbalanced operations. For example, the time consumed by the $\mathbf{K}\mathbf{W}^K$, $\mathbf{Q}\mathbf{W}^Q$ and $\mathbf{V}\mathbf{W}^V$ is roughly 4 times of computation required by the $n$ heads. To improve the utilization of pipeline, it is desirable to let each layers consumes roughly the same time. This can be achieved by allocating more resources to the slowest layer. We adopt a two-stage optimization flow. In first stage, we find a resource scheme that can minimize the maximum time required by layers. In second stage, under such resource constrains, we optimize the scheduling of an individual encoder/decoder.


The optimization starts from a basic implementation of an individual encoder/decoder, i.e. no parallization nor resource reusing, such that we can obtain an estimation of resource consumption, number of operations and execution time of each layer, throughput obtained by unit number of resources. Then we will examine how much resource can be allocated to each encoder/decoder to minimize the execution time of the slowest layer:

\vspace{-0.3cm}
\begin{equation}\label{eq:objective1}
\begin{aligned}
& {\textbf{minimize}} & & \max (T_1, T_2, ... , T_n),
\\ & {\textbf{subject to}}
& & R_{F}[i] \geq M \sum_j{ R_j[i]} + R_{misc}[i]
\end{aligned}
\end{equation}
\vspace{-0.3cm}

\noindent where $i \in (0,...,3)$, $j \in n$, $n$ is the number of layers, $M$ is the total number of encoder/decoder, $R_F = [R_{FF}, R_{LUT},R_{DSP}, R_{BRAM}]$ is on-chip resource constraints for look-up table (LUT), flip-flop (FF), digital signal processing unit (DSP), and BRAM, respectively.  
$T_j$ is the time required by the $j$-{th} layer. 
$R_j$
is resource utilization of the $j$-{th} layer, which is also represented as a vector: $R_j = [R^j_{FF}, R^j_{LUT},R^j_{DSP}, R^j_{BRAM}]$. $R_{misc}$ is the resource utilization of modules except encoder/decoder module, such as DDR controller, PCIE controller, etc.
$T_j$ can be given as:
\vspace{-0.3cm}
\begin{equation}
T_j = \ceil{{N^i_{op}}/(F_{j} \cdot K_j)}, j \in {n}
\vspace{-0.3cm}
\end{equation}
where $N^i_{op}$ is the number of operations required by the $j$-{th} layer. $K_{j}$ is resource allocation factor of the $j$-{th} layer. $F_j$ is the throughput of non-optimized design, which can be obtained empirically. Therefore, the throughput is:

\vspace{-0.3cm}
\begin{equation}
Throughput = freq /( n \cdot {max(T_1, T_2, ... , T_j)})
\end{equation}
\vspace{-0.5cm}

It finds the slowest layer, allocates more resources, then updates the resource consumption and execution time. If resource constraints are satisfied, we repeat this procedure until no more speedup. Then the algorithm will examine the fastest layer. If it takes significantly less time than the slowest layer, it is possible to allocate less resources for that layer, hence more resources can be assigned to the slowest layer. After this procedure, we obtain resource constraints, e.g. the No. of different PEs of an encoder and decoder.
 Under resource constraints, each layer may not have dedicated computation resource, hence matrix multipliers, adders, etc. have to be shared. Therefore, the computation has to be carefully scheduled to minimize the computation latency. The encoder/decoder can be represented as a Directed Acyclic Graph (DAG) -- ${G}({V}, {E})$, where ${V}$ is a set of vertices representing different computation, edges ${E}$ indicate the dependency. The available computation units such as PEs and adders are represented by a set ${Op} = \{PE-A1, PE-A2, ... , Adder\}$. The algorithm used for operation scheduling takes ${G}$ and ${Op}$ as input, is shown in Algorithm \ref{alg:1}.

\begin{algorithm}[t]
\scriptsize
\KwIn{Dependency graph ${G}({V}, {E})$, available PEs ${Op} = \{PE\_A1, PE\_A2, ... , Adder\}$}
 \KwOut{$C$ and designated $PE$ for each Layer}
 $ Q = TOPO\_SORT({G}({V}, {E}))$ $\backslash \backslash$ Topological sort to obtain priority queue of all layers \\
 $ P = Q[0]$ $\backslash \backslash$ List of layers to be scheduled \\
 $ E = \emptyset$ $\backslash \backslash$ List of layers being executed \\
 $ S = \emptyset$ $\backslash \backslash$ The final schedule result \\
 $stage = 0$ \\
 \While{$Q \neq \emptyset \land E \neq \emptyset$}
 {
    \For{$layer \in Q$}
    {
        \uIf{\textup{available} $PE \exists Op$ \textup{for} $layer$}
        {
            $Q.pop()$\\
            $Op.remove(PE)$ \\
            $E.push\_back( (layer, PE))$\\
            \For{$V \in NEIGHBOR(layer)$}
            {
            $Q.push\_back(V)$\\
            }
        }
    }
    $stage += 1$ \\
    \For{$layer, PE \in E$}
    {
        \uIf{$IS\_FINISHED(layer) == True$}
        {
            $E.pop()$\\
            $S.push\_back( (layer, stage, PE) )$\\
            $Op.push\_back(PE)$
        }
    }
 }

 \Return{$S$}
 \caption{Pseudo-code for operation scheduling}
 \label{alg:1}
\end{algorithm}

\section{Evaluation}

\subsection{Training of Transformer-Based Language Representation}

In this section, we apply both enhanced BCM-based model compression on the linear layers, and adopt 16 fixed-point data representation for all the weights. We 
evaluate the accuracy impact with two representative Transformer structures, i.e., a shallow Transformer with both encoder and decoder, and a pretrained deep Transformer architecture - RoBERTa (base configuration) which only has encoder~\cite{liu2019roberta}. 
The shallow Transformer is evaluated in a language modeling task, which is an unsupervised sequence-to-sequence problem that requires the decoder part. On the other hand, we run a RoBERTa on a sentiment classification task that is a supervised classification problem without the requirement for decoder block. 
The software is implemented in PyTorch deep learning framework \cite{paszke2017automatic} and FairSeq sequence modeling toolkit \cite{ott2019fairseq}. 
Table \ref{tbl:size} summarizes the key parameters of the shallow Transformer and RoBERTa models in the experiments.


\begin{table}[h]
\caption{Key parameters of Shallow Transformer and RoBERTa}\label{tbl:size}
\begin{center}
\resizebox{0.95\columnwidth}{!}{
\begin{tabular}{c|c|c|c|c|c}
\hline
Model & Transformer &Transformer  & Hidden  & Attention  & Total  \\
Configuration & Structure &Layers & Size & Heads & Params
\\\hline 
Shallow Transformer & encoder-decoder & 2 &  200 & 4 & 6M  \\
RoBERTa (base config.) & encoder only & 12 & 768 & 12 & 125M \\\hline
\end{tabular}}
\end{center}
\end{table}


\subsubsection{Finetuned RoBERTa for Sentiment Classification}
We evaluate the proposed model compression method for finetuned RoBERTa~\cite{liu2019roberta} on IMDB movie review sentiment classification \cite{maas-EtAl:2011:ACL-HLT2011} to shed some light on training trial reductions. 
Starting from the saved state of pretrained models in work~\cite{liu2019roberta}, we finetune the model until it reaches to its best validation accuracy at 95.7\%. To maintain overall accuracy, we compress partial layers.
The process suppresses randomness by using a deterministic seed. Thus the accuracy difference between the original RoBERTa and compressed version is sorely contributed by the compression techniques.

\subsubsection{Shallow Transformer}
Language modeling task takes a sequence of words as input and determines how likely that sequence is the actual human language.
We consider the popular WikiText-2 dataset \cite{merity2016pointer} in this experiment, which contains 2M training tokens with a vocabulary size of 33k. A shallow Transformer model with 4 attention heads and 200 hidden dimension is established. 

The baseline and model compression results of shallow Transformer and RoBERTa on WikiText-2 and IMDB review are shown in Table \ref{tbl:GRU_accuracy}, respectively. We compress the models using enhanced BCM-based method with block size of 4 or 8. 
From Table \ref{tbl:GRU_accuracy}, we observe that for the shallow Transformer, thers is no accuracy loss with block size of 4 and only 0.6\% accuracy loss with block size of 8. 
The RoBERTa, on the other hand, incurs 4.2\% and 4.3\% accuracy drop after model compression using 4 and 8 block size, respectively\footnote{The accuracy drop on RoBERTa is slightly higher because its parameters are carefully pretrained on the Giga byte dataset (160GB of text) using a masked language model~\cite{liu2019roberta} and more sensitive to compression. }. 
We also observe that changing from 32-bit floating point to 16-bit fixed point will not cause accuracy loss.
The comparable accuracy between the original model and the weight compressed version demonstrates the effectiveness of the proposed model compression method.





\begin{table}[t]
	\centering
	\caption{Comparison among different model configurations}
	\vskip -0.8em
	\label{tbl:GRU_accuracy}
	\resizebox{\columnwidth}{!}{
		\begin{tabular}{c|c|c|c|c|c}
			\hline
			\multirow{2}{*}{ID}  & Network &Block  & WikiText-2  & ACC loss& ACC loss\\
		 	& Type  & Size                         & (ACC) \% & with BCM  (\%)& with BCM \& Quant.  (\%)\\\hline
			1 &Shallow Transformer       & $-$  &91.3 &  $-$ & $-$\\\hline
			2 &Shallow Transformer       & $4$  & 90.7& 0.6  & 0\\\hline
			3 &Shallow Transformer       & $8$  & 90.7& 0.6  & 0.6\\\hline
			4 &Shallow Transformer       & $16$  & 90.0&  1.3 & 0.6\\\hline\hline
			\multirow{2}{*}{ID}  & Network &Block  & IMDB  & ACC loss& ACC loss\\
		 	& Type  & Size                         & (ACC)\% & with BCM  (\%)& with BCM \& Quant.  (\%)\\\hline
			4 &RoBERTa (base)        & $-$    & 95.7&  $-$ &$-$\\\hline
			5 &RoBERTa (base)       & $4$  & 91.5& 4.2 & 4.3\\\hline
			6 &RoBERTa (base)          & $8$  & 91.4& 4.3 &4.3 \\\hline\hline
		\end{tabular}
	}
\end{table}

\subsection{Performance and Energy Efficiency}

\begin{table}[t]
\centering
\caption{Comparison among different model configurations}
\label{table:tbl2}
\resizebox{\columnwidth}{!}{
\begin{tabular}{l|l|l|l|l|l|l}
\hline
\multicolumn{7}{c}{Shallow Transformer}                              \\ \hline
Batch Size & DSP  & FF     & LUT    & Latency (ms) & Power (W)  & Throughput (FPS) \\ \hline
1          & 5647 & 304012 & 268933 & 2.94    & 22.45  &   680.91         \\ \hline
4          & 5647 & 304296 & 269361 & 11.59   & 22.52  &   690.50        \\ \hline
8          & 5647 & 305820 & 269753 & 22.90   & 22.66  &   698.72         \\ \hline
16         & 5647 & 306176 & 270449 & 45.54   & 22.73  &   702.54        \\ \hline \hline
\multicolumn{7}{c}{RoBERTa (base)}                                  \\ \hline
Batch Size & DSP  & FF     & LUT    & Latency (ms) & Power (W)  & Throughput (FPS) \\ \hline
1          & 6531 & 506612 & 451073 & 10.61   & 25.06 &     94.25       \\ \hline
4          & 6531 & 506936 & 451545 & 40.33   & 25.13 &     99.13       \\ \hline
8          & 6531 & 508488 & 452005 & 79.03   & 25.89 &     101.23      \\ \hline
16         & 6531 & 508916 & 452661 & 157.18  & 25.96 &     101.79      \\ \hline
\end{tabular}
}
\end{table}


 \subsubsection{Experimental Platform}
The Xilinx Virtex UltraScale+ VCU118 board, comprising 345.9Mb BRAM, 6,840 DSPs, 2,586K logic cells (LUT), and  two 4GB DDR5 memory, is connected to the host machine through PCIE Gen3 $\times$ 8 I/O Interface. The host machine adopted in our experiments is a server configured with multiple Intel Core i7-8700 processors. 
We use {Xilinx SDX 2017.1} as the commercial high-level synthesis backend to synthesize the high-level (C/C++) based designs on the selected FPGAs. 

\subsubsection{Experimental Results of Transformer and RoBERTa}
We implement the compressed model to FPGA to evaluate the performance and energy efficiency. For different batch sizes, we obtain the parallelism per stage for the 7 stages in encoder/decoders of Transformer and  RoBERTa based on Algorithm~\ref{alg:1} as shown in Table~\ref{table:tbl2}, respectively. We report the resource utilization on FPGA including DSP, LUT, and FF. The latency (ms), throughput (frame/sequence per second) and power consumption (W) are also reported. Our results shows that there is a trade-off between latency and power consumption. For both Transformer and RoBERTa, we can achieve the best trade-off (the lowest ratio of Latency/Power) when the batch size is 8 since the latency will be significantly increased and the throughput will not be increased when we use larger batch size. 

\subsubsection{Cross-platform comparison}
We compare the performance (throughput) and energy efficiency among CPU, GPU, and FPGA using same model and same benchmark (IMDB), as shown in Table~\ref{table:tbl3}. 
We also validate our method on embedded low-power devices, implement our pruned model on Jetson TX2, an embedded AI computing device. It’s built by a 256-core NVIDIA Pascal-family GPU and the memory is 8 GB with 59.7 GB/s bandwidth.  Our FPGA design achieves 27.07$\times$ and 81$\times$ improvement in throughput and energy efficiency compared to CPU. For GPU TRX5000, the power consumption is 5.01$\times$ compared to that of FPGA, and Our FPGA design achieves 8.80$\times$ improvement in energy efficiency and 1.77$\times$ throughput improvement compared to GPU. For embedded GPU Jason TX2, our FPGA design achieves 2.44$\times$ improvement in energy efficiency.

\begin{table}[!ht]
    \centering
    \caption{The performance and energy efficiency comparison among CPU, GPU, FPGA using RoBERTa}\vspace{-0.1cm}\label{table:tbl3}
    \resizebox{0.9\columnwidth}{!}{
        \begin{tabular}{c|c|c|c|c}
         \hline
            & CPU & GPU & FPGA & Jetson TX2\\
            &  i7-8700K & RTX5000 & VCU118 &  Embedded GPU\\        
             \hline
            Throughput (FPS) & 3.76 & 57.46 & 101.79 & 9.75\\
            \hline
            Power (W) &80 & 126 &  25.13 & 5.86\\
            \hline
            Energy efficiency  (FPS/W) &0.05 & 0.46& 4.05  & 1.66\\ 
        \hline
        \end{tabular}
    }
\end{table}

\section{Conclusion}

In this paper, we propose an energy-efficient acceleration framework for transformer-based large scale language representations. Our framework includes an enhanced BCM-based method to enable model compression on large-scale language representations at the algorithm level, and an acceleration design at the architecture level. We propose an FPGA architecture design to support the model compression technique and we develop a design automation and optimization technique to explore the parallelism and achieve high throughput and performance.
Experimental results show that our proposed framework significantly reduces the size of NLP models with small accuracy loss on Transformer. Our FPGA-based implementation significantly outperforms CPU and GPU in terms of energy efficiency. 


\bibliographystyle{ACM-Reference-Format}
\bibliography{bib}


\end{document}